\documentclass[12pt]{article}
\usepackage{putex}
\usepackage{graphicx}
\usepackage{latexsym,amsmath,amsfonts,amssymb,cite}
\usepackage{bbm}
\numberwithin{equation}{section}

\def\btab{\begin{table}[h] \begin{center} \begin{tabular}{l lp{3in}}}
      \def\etab{\end{tabular} \end{center} \end{table}}
\def\btabm{\begin{center} \begin{tabular}}
    \def\etabm{\end{tabular} \end{center}}

\def\ie{{\it i.e.}}

\def\D{{\Delta}}

\def\CR{{\cal R}}

\def\CN{{\cal N}}

\def\CP{{\cal P}}

\def\CW{{\cal W}}

\def\BR{\mathbb{R}}

\begin{document}
%
\title{The Gravity Dual of Supersymmetric R{\'e}nyi Entropy}

\authors{Tatsuma Nishioka}

\institution{??}{
School of Natural Sciences, Institute for Advanced Study, Princeton, NJ 08540, USA}

\abstract{
Supersymmetric R{\' e}nyi entropies are defined for three-dimensional $\CN=2$ superconformal field theories on a branched covering of a three-sphere by using the localized partition functions.
Under a conformal transformation, the branched covering is mapped to $S^1 \times H^2$, whose gravity dual is the charged topological AdS$_4$ black hole.
The black hole can be embedded into four-dimensional $\CN=2$ gauged supergravity where the mass and charge are related so that it preserves half of the supersymmetries.
We compute the supersymmetric R{\' e}nyi entropies with and without a certain type of Wilson loop operators in the gravity theory.
We find they agree with those of the dual field theories in the large-$N$ limit.
}

\date{January 2014}
\maketitle

\tableofcontents
%

\section{Introduction}
In this paper, we consider the gravity duals of the supersymmetric R{\' e}nyi entropies  for three-dimensional $\CN=2$ superconformal field theories, which are defined by \cite{Nishioka:2013haa}
\begin{align}\label{SRE}
S^\text{susy}_n = \frac{1}{1-n} \log \left| \frac{Z_n}{(Z_1)^n}\right| \ .
\end{align}
Here $Z_n$ is the supersymmetric partition function on a branched $n$-covering of a three-sphere 
\begin{align}\label{ConicalS3}
\begin{aligned}
ds^2 &= d\theta^2 + \sin^2 \theta d\tau^2 + \cos^2 \theta d\phi^2 \ , \\
&\theta \in [0, \pi/2] \ , \qquad \tau \in [0, 2\pi n ) \ , \qquad \phi \in [0, 2\pi ) \ .
\end{aligned}
\end{align}
The supersymmetric R{\' e}nyi entropies differ from the usual R{\' e}nyi entropies of a disk entangling region in $\BR^{2,1}$ due to the nontrivial $R$-symmetry background gauge field for preserving half of the supersymmetries. 
Therefore, the existing results of the holographic R{\' e}nyi entropies \cite{Hung:2011nu} can not be applied to them.\footnote{Actually, the supersymmetric R{\' e}nyi entropies are parts of charged R{\' e}nyi entropies defined in \cite{Belin:2013uta}.}

\begin{figure}[htbp]
\centering
\includegraphics[width=12cm]{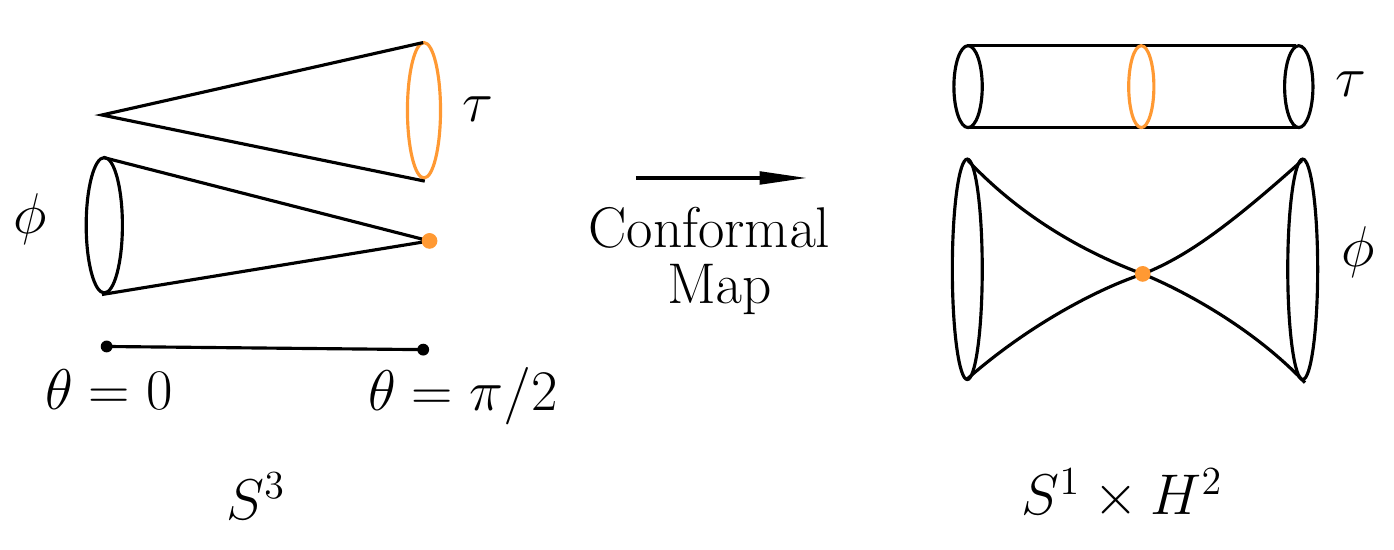}
\caption{The conformal map from the branched covering of a three-sphere \eqref{ConicalS3} to $S^1 \times H^2$ \eqref{Hyp}.
The orange circle at $\theta=\pi/2$ in the left figure is mapped to a circle (also shown in orange color) along $\tau$ direction at $u=0$ of the hyperbolic space $H^2$ in the right figure.
}
\label{fig:ConformalMap}
\end{figure}

To find the gravity duals, we will look for a half-BPS solution in four-dimensional $\CN=2$ gauged supergravities. 
Instead of dealing with the branched $n$-covering three-sphere \eqref{ConicalS3}, 
one finds it easier to conformally map it by $\cot \theta = \sinh u$ to the spacetime $S^1\times H^2$ \cite{Casini:2011kv}
\begin{align}\label{Hyp}
\begin{aligned}
ds^2 &= d\tau^2 + du^2 + \sinh^2 u \,d\phi^2 \ , \\
&\tau \in [0, 2\pi n ) \ , \qquad u\in [0, \infty) \ , \qquad \phi \in [0, 2\pi ) \ .
\end{aligned}
\end{align}
The advantage of these coordinates is that the conical singularity, which was at $\theta = 0$ in \eqref{ConicalS3}, is pushed away to the spatial infinity $u\to \infty$ (see Fig.\,\ref{fig:ConformalMap}).
The AdS/CFT dictionary \cite{Maldacena:1997re,Gubser:1998bc,Witten:1998qj} implies that the dual gravity theory should have $U(1)$ gauge symmetry corresponding to the $R$-symmetry background gauge field of the field theory.
Thus we are led to consider the charged AdS$_4$ topological black holes with half supersymmetries \cite{AlonsoAlberca:2000cs} to construct
holographic duals of the supersymmetric R{\' e}nyi entropies.

The organization of this paper is as follows. 
In section \ref{ss:SRE}, we start with the supersymmetric R{\' e}nyi entropies of three-dimensional $\CN=2$ superconformal field theories.
The partition function $Z_n$ is given by the matrix model on a squashed three-sphere $S^3_b$ with the squashing parameter $b=\sqrt{n}$.
We calculate $Z_n$ for a class of theories whose gravity duals are described by M-theory in the large-$N$ limit.
Then we add a Wilson loop in a fundamental representation and estimate the shift of the supersymmetric R{\' e}nyi entropies for the ABJM theory \cite{Aharony:2008ug} in the large-$N$ limit.
Section \ref{ss:HSRE} is concerned with the gravity duals.
The partition function in \eqref{SRE} is given by the Euclidean on-shell action $I(n)$ in the dual gravity theory, $Z_n = e^{-I(n)}$, of the charged topological AdS$_4$ black holes at finite temperature $T =1/(2\pi n)$.
The on-shell action is evaluated in a standard way.
As a gravity dual of the Wilson loop, we consider the fundamental string and calculate the supersymmetric R{\' e}nyi entropies.
We find the holographic computations agree with the field theory results in the large-$N$ limit.

\bigskip
While this paper was in preparation, \cite{Huang:2014gca} appeared, which has a substantial overlap with this paper.

\section{Supersymmetric R{\'e}nyi entropy}\label{ss:SRE}

The partition function $Z_n$ was calculated by localization method in \cite{Nishioka:2013haa}.
It turns out to be the partition function on a squashed three-sphere $S^3_b$ \cite{Hama:2011ea,Imamura:2011wg} with the squashing parameter $b=\sqrt{n}$:
\begin{align}\label{MM}
\begin{aligned}
Z_{n} = &\frac{1}{|\CW|}\int\prod_{i=1}^{\text{rank}\, G}d\sigma_{i}\, e^{-F_n} \ , \\
F_n = & - \pi ik\,\text{Tr}(\sigma^{2})+ \sum_{\alpha}\log\Gamma_{h}\left(\alpha(\sigma)\right) - \sum_{I}\sum_{\rho\in\CR_{I}}\log\Gamma_{h}\left(\rho(\sigma)+i\omega\D_{I}\right)\ ,
\end{aligned}
\end{align}
where $\sigma_i$ are the eigenvalues of the matrix $\sigma$ and 
\begin{align}
\Gamma_{h}\left(z\right)\equiv\Gamma_{h}\left(z;i\omega_{1},i\omega_{2}\right)\ ,\qquad\omega_{1}=\omega_2^{-1} = \sqrt{n}\ ,\qquad \omega=\frac{\omega_{1}+\omega_{2}}{2}\ ,
\end{align}
is the hyperbolic gamma function,\footnote{The hyperbolic gamma function is a meromorphic function of a single
complex variable with two parameters defined in \cite{van2007hyperbolic}
\begin{align}
\begin{aligned}\Gamma_{h}\left(z;\omega_{1},\omega_{2}\right) & =\prod_{n_{1},n_{2}\ge0}\frac{(n_{1}+1)\omega_{1}+(n_{2}+1)\omega_{2}-z}{n_{1}\omega_{1}+n_{2}\omega_{2}+z} \ ,
\end{aligned}
\end{align}
with the integral defined for 
$0<\Im\left(z\right)<\Im\left(\omega_{1}+\omega_{2}\right)$ 
and then analytically continued to the entire complex plain. } $|\CW|$ is the order of the Weyl group $\CW$ of the gauge group $G$ and
$k$ is the Chern-Simons level.\footnote{We can turn on the Fayet-Iliopoulos and real mass terms, but we will not do so here.}
$I$ labels the
types of chiral multiplets and $\rho$ is a weight in a representation
$\CR_{I}$ of the gauge group $G$. $\D_{I}$ is the $R$-charge of
the scalar field in a chiral multiplet.

Performing the integration of the matrix model \eqref{MM} analytically is limited for simple theories, while its large-$N$ limit becomes tractable as we will see below.

\subsection{Large-$N$ limit}
We restrict our attention to a class of quiver Chern-Simons gauge theories which are dual to M-theory in the large-$N$ limit.
Namely, we consider non-chiral gauge theories with the gauge group $G=U(N)_{k_1} \times U(N)_{k_2} \times \cdots \times U(N)_{k_p}$ and bifundamental fields.
The large-$N$ analysis on a round three-sphere \cite{Herzog:2010hf,Martelli:2011qj,Cheon:2011vi,Jafferis:2011zi} shows the theories must satisfy $\sum_{a=1}^p k_a = 0$ and $\sum_{I\in a} (1-\D_I) = 2$ for every gauge group labeled by $a=1,\cdots, p$ and bifundamental fields of representation $I$ of the $R$-charge $\Delta_I$. 
We proceed to assume these conditions for taking the large-$N$ limit of the partition function \eqref{MM}.
The result has already been obtained in \cite{Imamura:2011wg,Martelli:2011fu}, and we briefly review it for later use. 

For each unitary group $U(N)_{k_a}, a=1,\cdots, p$, there are eigenvalues $\sigma_{a,i}$ with $i=1,\cdots, N$ with respect to which the free energy $F_n$ in \eqref{MM} is extremized, $\partial F_n/ \sigma_{a,i} =0$.
To solve the saddle-point equations,
we take the large-$N$ limit by introducing the eigenvalue density $\rho(x) = \frac{1}{N}\sum_{i=1}^N \delta (x-x_i)$ and assuming the eigenvalues lie along 
\begin{align}
\sigma_{a,i} = N^{1/2} x_{i} + i \omega\, y_{a,i} \ .
\end{align}
The $\omega$ in front of $y_{a,i}$ is put for later convenience.
This assumption yields a consistent solution to the saddle-point equations.
The point is the separation of the long and short range forces between the eigenvalues.
We employ the approximation of the hyperbolic gamma function with $\omega_1 = \omega_2^{-1}$ \cite{Imamura:2011wg}\footnote{In \cite{Imamura:2011wg}, the approximation of the double sine function $s_b(z)$ is obtained, which is related to the hyperbolic gamma function as $\Gamma_h (z) = s_b (i\omega -z)$ for $\omega_1 = \omega_2^{-1}= b$.}
\begin{align}
\begin{aligned}
\log \Gamma_h (z) \simeq &- \frac{\pi i}{2} \left[ (z-i \omega)^2 + \frac{\omega^2}{3} - \frac{1}{6}\right]\text{sgn}(\text{Re}\, z)   \\ 
	& + \frac{\pi}{3} (\text{Im}\,z -\omega) \left( (\text{Im}\,z)^2 - 2\omega \,\text{Im}\,z + \frac{1}{2}\right) \delta (\text{Re}\,z) \ .
\end{aligned}
\end{align}
With this form, the long range force vanishes and the free energy is proportional to $N^{3/2}$ 
\begin{align}
\begin{aligned}
F_n/N^{3/2} &= 2\pi \omega \sum_{a=1}^p k_a \int dx\, \rho(x) \, x\,  y_a (x) \\
	& + \sum_{I\in\text{bi-fund}}\frac{\omega^3}{3}   \int dx\, \rho(x)^2 (y_I (x) + \D_I) (y_I(x) + \D_I -1)(y_I(x) + \D_I -2) \ ,
\end{aligned}
\end{align}
where $y_I (x) = y_a(x) - y_b(x)$ for the indices of the gauge groups $a,b$ the bifundamental field $I$ belongs.

The first line is proportional to $\omega$ while the second is to $\omega^3$. 
If one rescales the Chern-Simons level by $k_a = \omega^2 \hat k_a$, the whole term becomes $\omega^3$ times the free energy for the theory with the levels $\hat k_a$.
Since we know the free energy on $S^3$ (\ie, $n=1$) of our gauge theories dual to M-theory is proportional to the square root of the levels \cite{Martelli:2011fu}, we obtain the relation
\begin{align}\label{Fn}
F_n(N, k_a) = \omega^3 F_1 (N, \hat k_a) = \omega^2 F_1 (N, k_a) \ ,
\end{align}
where we explicitly write the dependence of the free energy on the parameters.
Substituting into the definition \eqref{SRE}, the supersymmetric R{\'e}nyi entropy in the large-$N$ limit becomes
\begin{align}\label{SRELarge}
	S_n^\text{susy} = - \frac{3n + 1}{4n} F_1\ .
\end{align}

Finally, we quote the result of the eigenvalue density of the ABJM theory with the gauge group $G=U(N)_k\times U(N)_{-k}$ from \cite{Herzog:2010hf}.
It is a constant function on a compact support 
\begin{align}\label{ABJM}
\rho_\text{ABJM}(x) = \frac{1}{2\pi}\sqrt{\frac{k}{2}} \ , \qquad x\in \left[- \pi \sqrt{\frac{2}{k}}, \pi \sqrt{\frac{2}{k}}\right] \ .
\end{align}
Using the eigenvalue density, the supersymmetric R{\'e}nyi entropy for the ABJM theory in the large-$N$ limit is given by
\begin{align}
S_n^\text{susy} = - \frac{3n + 1}{4n}\frac{\pi \sqrt{2}}{3}k^{1/2} N^{3/2} \ .
\end{align}

\subsection{Wilson loop}
Next, we consider a Wilson loop at $\theta =\pi/2$ wrapping around $\tau$ direction of the branched $n$-covering three-sphere \eqref{ConicalS3} (see Fig.\,\ref{fig:ConformalMap}). 
After the conformal mapping, this is equivalent to a temporal Wilson loop along $\tau$
at $u=0$ in CFT on $S^1\times H^2$ given in \eqref{Hyp}.
The addition of the Wilson loop $W$ shifts the entanglement entropy by \cite{Lewkowycz:2013laa}
\begin{align}
S_{W} = \lim_{n\to 1} (1- \partial_n)\log |\langle W\rangle_n|  \ ,
\end{align}
where $\langle \cdot \rangle_n$ is the expectation value on the branched $n$-covering three-sphere.
Similarly, the shift of the R{\' e}nyi entropy due to the loop takes the form
\begin{align}\label{SREWilson}
S_{W,n} = \frac{1}{n-1}\left( n \log |\langle W\rangle_1| - \log |\langle W\rangle_n| \right) \ .
\end{align}

For $\CN =2$ Chern-Simons gauge theories, the supersymmetric Wilson loop in representation $R$ is
\begin{align}
W = \text{Tr}_R\, \CP \exp\left[ \oint ds\left( i A_\mu \dot x^\mu(s) + \sigma |\dot x(s)|\right)\right] \ ,
\end{align}
where $x^\mu(s)$ is the location of the Wilson loop \cite{Gaiotto:2007qi,Kapustin:2009kz}.
The expectation value of the Wilson loop on the $n$-covering three-sphere can be obtained by localization following \cite{Nishioka:2013haa}, and reduces to the matrix model \cite{Lewkowycz:2013laa}
\begin{align}
\langle W\rangle_n = \frac{1}{Z_n |\CW|}\int\prod_{i=1}^{\text{rank}\, G}d\sigma_{i}\, \text{Tr}_R (e^{2\pi \omega_1 \sigma}) \, e^{-F_n} \ ,
\end{align}
where we used the notations in \eqref{MM}.

In the large-$N$ limit, we can represent the expectation value of the Wilson loop in the fundamental representation using the eigenvalue density $\rho(x)$ as 
\begin{align}
 \langle W\rangle_n = \int dx\, \rho(x)\, e^{2\pi \omega_1 \left(N^{1/2}x + O(1)\right)}\ .
\end{align}
Let us apply this formula to the $1/6$-BPS Wilson loop in the ABJM theory. The eigenvalue density is given in \eqref{ABJM}, but we need to scale the Chern-Simons level by $\omega^2$ on the $n$-covering three-sphere as we mentioned above \eqref{Fn}, resulting in
\begin{align}\label{LogW}
\log  \langle W\rangle_n = \frac{\pi (n+1)}{2} \sqrt{2\lambda}  + O(\log N)\ ,
\end{align}
where $\lambda$ is the 't Hooft coupling $\lambda=N/k$. 
Thus combining with \eqref{SREWilson}, we obtain the supersymmetric R{\'e}nyi entropy of the Wilson loop of the ABJM theory in the large-$N$ limit
\begin{align}\label{SREW}
S_{W,n} =  \frac{\pi}{2}\sqrt{2\lambda} \ .
\end{align}
It coincides with the result of \cite{Lewkowycz:2013laa} for $n=1$.
Moreover, it does not depend on the parameter $n$.
In the next section, we will consider the gravity duals of the supersymmetric R{\'e}nyi entropies with(out) a Wilson loop and see if they agree with the results in this section.

\section{The charged topological AdS black hole}\label{ss:HSRE}
In this section, we construct the gravity duals of the supersymmetric R{\'e}nyi entropies.
Instead of finding a solution whose boundary is the branched covering of three-sphere \eqref{ConicalS3}, we will look for a solution which asymptotes to $S^1 \times H^2$ \eqref{Hyp} near the boundary.
Since we turn on the $U(1)$ $R$-symmetry background gauge field for the boundary SCFT to preserve half of the supersymmetries, there is a $U(1)$ gauge field in the corresponding bulk theory.

We consider the Einstein-Maxwell theory in (Euclidean) four-dimensions
\begin{align}\label{EuclideanAction}
I = - \frac{1}{16\pi G_4}\int d^4 x \sqrt{g}\left[ R + 6g^2 - F_{\mu\nu}F^{\mu\nu}\right] \ .
\end{align}
There exists a charged topological AdS black hole solution whose boundary is $S^1\times H^2$ \cite{Brill:1997mf
}
\begin{align}\label{CTADS}
ds^2 = \frac{f(r)}{g^2} d\tau^2 + \frac{1}{f(r)} dr^2 + r^2 (du^2 + \sinh^2 u\, d\phi^2) \ ,
\end{align}
where 
\begin{align}\label{lapse}
f(r) =g^2 r^2 -1 + \frac{2M}{r} - \frac{Q^2}{r^2}  \ ,
\end{align}
and the $U(1)$ gauge field takes the form
\begin{align}
A_\mu dx^\mu =\left( \frac{Q}{g r} - \mu \right) d\tau \ .
\end{align}
Let $r_H$ be the largest root of $f(r)$. 
The chemical potential $\mu$ can be determined by demanding that $A_\mu$ vanishes on the horizon at $r=r_H$:
\begin{align}
\mu = \frac{Q}{g r_H} \ .
\end{align}

The theory \eqref{EuclideanAction} can be embedded into the $\CN=2$ gauged supergravity \cite{Freedman:1976aw} where the Killing spinor equation arises as a variation of the gravitino
\begin{align}
\left[ \nabla_\mu + \frac{g}{2} \Gamma_\mu - ig A_\mu + \frac{i}{4}F_{\nu\rho}\Gamma^{\nu\rho} \Gamma_\mu \right] \zeta = 0 \ .
\end{align}
Here
$\Gamma_a$ are the gamma matrices for the local Lorentz coordinates satisfying $\{ \Gamma_a, \Gamma_b\} = 2\delta_{ab}$.
$\Gamma_\mu = \Gamma_a e^a_\mu$ is the pullback by the vielbein $ds^2 = \sum_{a=1}^4 (e^a)^2$. 
The above solution \eqref{CTADS} preserves $1/2$ supersymmetries when the following condition holds \cite{AlonsoAlberca:2000cs}:
\begin{align}
M = Q \ .
\end{align}
We will only consider the supersymmetric solution. 
Then $Q$ is given by
\begin{align}
Q = r_H (g\, r_H - 1) \ ,
\end{align}
and the temperature of the black hole is
\begin{align}\label{Temperature}
T =  \frac{2g\, r_H - 1}{2\pi}\ , 
\end{align}
The thermal entropy is
\begin{align}
S_\text{therm} = \frac{\text{Vol}(H^2)}{4G_4}r_H^2 \ .
\end{align}

To calculate the free energy, we evaluate the action \eqref{EuclideanAction} of the solution \eqref{CTADS} with the boundary and counter terms \cite{Emparan:1999pm,Gabella:2011sg}
\begin{align}
\begin{aligned}
I_\text{bdy} &= - \frac{1}{8\pi G_4}\int_B d^3 x\sqrt{\gamma} \, K \ , \\
I_{ct} &= \frac{1}{8\pi G_4}\int_Bd^3 x\sqrt{\gamma} \left[ 2g + \frac{1}{2g}R_B \right] \ ,
\end{aligned}
\end{align}
where $\gamma_{\alpha\beta}$ is the induced metric on the boundary $B$, $K$ the extrinsic curvature, and $R_B$ the Ricci scalar of the induced metric.
These terms are needed to regularize the UV divergence coming from the infinite volume of the boundary $r=r_\infty \to \infty$.
After a little computation, we obtain the free energy of the supersymmetric charged topological AdS$_4$ black hole
\begin{align}\label{SAdSTop}
\begin{aligned}
I_\text{tot} &= I + I_\text{bdy} + I_{ct} \ , \\
	&= \frac{\text{Vol}(H^2)}{8\pi G_4} \frac{g}{T}\left( -r_H^3 + \frac{Q}{g^2} + \frac{Q^2}{g^2 r_H} \right) \ .
\end{aligned}
\end{align}

\subsection{Holographic supersymmetric R{\'e}nyi entropy}

We would like to compare the free energy with the result in the dual field theory \cite{Nishioka:2013haa}.
The temperature $T$ given by \eqref{Temperature} should be identified with $T_n = 1/(2\pi n)$, which determines the radius of the horizon
\begin{align}
r_H = \frac{n+1}{2g n} \ .
\end{align}
After all, we obtain the holographic free energy from \eqref{SAdSTop}\footnote{We use the regularized volume for $H^2$: Vol$(H^2)=-2\pi$.}
\begin{align}
F_n\equiv I_\text{tot} = \frac{(n+1)^2}{4n} \frac{\pi}{2G_4 g^2} \ ,
\end{align}
which satisfies the same relation to the large-$N$ limit of the free energy of the dual field theory \eqref{Fn}.
It follows that the holographic supersymmetric R{\'e}nyi entropy also agrees with \eqref{SRELarge}.

On the other hand, the usual holographic R{\'e}nyi entropy can be written as an integration of the thermal entropy \cite{Hung:2011nu,Belin:2013uta}
\begin{align}\label{HREformula}
S_n = \frac{n}{n-1}\frac{1}{T_1} \int_{T_n}^{T_1} S_\text{therm}(T) dT \ ,
\end{align}
which yields
\begin{align}\label{NaiveRE}
S_n = -\frac{(7n^2 + 4n + 1)\pi}{12n^2} F_1 \ .
\end{align}
The discrepancy between \eqref{SRELarge} and \eqref{NaiveRE} arises from the fact that there is the electric charge $Q$ and the chemical potential $\mu$ which depends on the temperature.
Namely, in deriving the formula \eqref{HREformula}, the thermodynamical relation $S_\text{therm}  = - \partial (I_\text{tot}/T)/\partial T|_{\mu: \,\text{fix}}$ was used \cite{Hung:2011nu}.

\subsection{Holographic Wilson loop}
A Wilson loop in a fundamental representation is holographically dual to the fundamental string in the space-time \eqref{CTADS} \cite{Maldacena:1998im,Rey:1998ik}.
\begin{figure}[htbp]
\centering
\includegraphics[width=10cm]{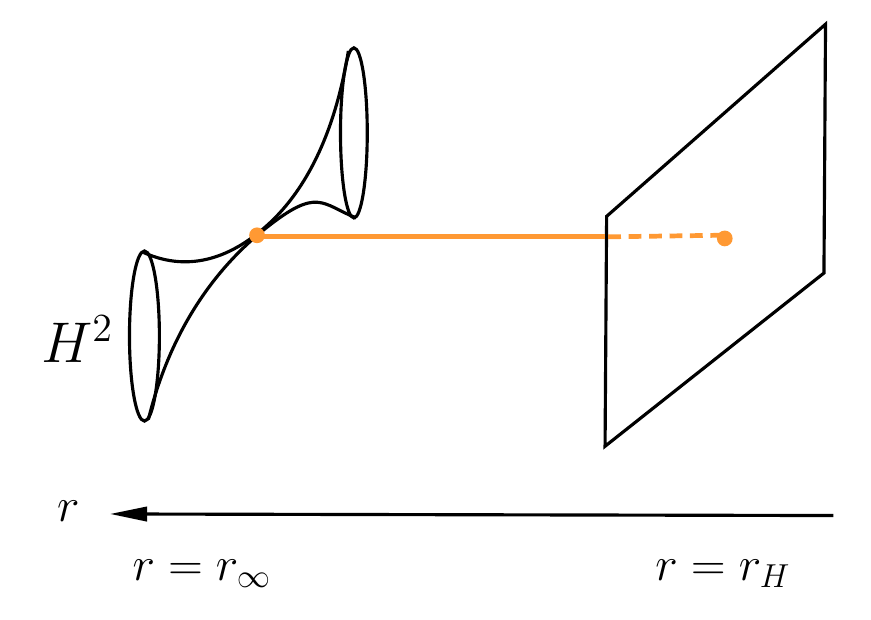}
\caption{The gravity dual of a Wilson loop in a fundamental representation is given by the fundamental string in AdS$_4$ spacetime.
In our case, it has two ends at the UV boundary $r=r_\infty$ and the horizon $r=r_H$. 
We suppress the $\tau$ direction in this figure.
}
\label{fig:Fstring}
\end{figure}
The string world sheet action with a target space $ds^2 = G_{\mu\nu} dx^\mu dx^\nu$
\begin{align}
S_\text{string} = \frac{1}{2\pi \alpha'} \int d^2\xi \sqrt{\det G_{\mu\nu}\partial_\alpha x^\mu \partial_\beta x^\nu} \ ,
\end{align}
leads to the expectation value 
\begin{align}
\log \langle W\rangle = -S_\text{string} \ .
\end{align}
If we identify the world sheet coordinates with the target space as $\xi^1 = \tau, \xi^2=r$, the rest of the target space coordinates do not depend on $\xi^\alpha$ because of the rotational symmetry of the Wilson loop (see Fig.\,\ref{fig:Fstring}).
Therefore, the on-shell action of the fundamental string on the background \eqref{CTADS} becomes
\begin{align}
S_\text{string} = \frac{n}{g^2\alpha'} (r_\infty - r_H) 
= - \frac{n+1}{2\alpha' g^2} \ ,
\end{align}
where we regulated the UV divergence near the boundary by the UV cutoff at $r=r_\infty$.
For a $1/6$-BPS Wilson loop in the ABJM theory with $U(N)_k\times U(N)_{-k}$ gauge group \cite{Drukker:2008zx,Chen:2008bp,Rey:2008bh}, we can rewrite the above result in terms of the 't Hooft coupling $\lambda = N/k = 1/(2\pi^2 \alpha'^2 g^4)$ and obtain
\begin{align}
\log \langle W\rangle_n = \frac{\pi (n+1)}{2} \sqrt{2\lambda} \ .
\end{align}
This agrees perfectly with the field theory computation \eqref{LogW} in the large-$N$ limit.
Again, it follows that the holographic supersymmetric R{\'e}nyi entropy of the Wilson loop agrees with \eqref{SREW}.

 \vspace{1.3cm}
 \centerline{\bf Acknowledgements}
We are grateful to J.\,Maldacena, T.\,Ugajin, I.\,Yaakov and D.\,Yokoyama for valuable discussions.
We also thank KIAS and 18th APCTP Winter School on Fundamental Physics for hospitality during the completion of the paper.
This work was supported by a JSPS postdoctoral fellowship for research abroad.


\bibliographystyle{JHEP}
\bibliography{Holographic_SuperRenyi}

\end{document}